\input{aipcheck}

\documentclass[
    ,final            % use final for the camera ready runs
  ]
  {aipproc}

\layoutstyle{8x11single}

\begin{document}

\title{Resolving the Axial Mass Anomaly in $\nu_{\mu}$  Scattering}


\author{A. Bodek}{
  address={Department of Physics and Astronomy, University of
Rochester, Rochester, NY  14627-0171 USA}
}

\author{H.S. Budd}{
  address={Department of Physics and Astronomy, University of
Rochester, Rochester, NY  14627-0171 USA}
}

\author{M. E. Christy}{
  address={Hampton University; Hampton, Virginia, 23668 USA}
  % ,altaddress={<author1 address>
  } % additional visiting address

\begin{abstract}
 We present a parametrization of  the observed  enhancement in the transverse electron  quasielastic (QE) response function for nucleons bound in carbon as a function of the square of the four momentum transfer ($Q^2$) in terms of a  correction to the magnetic form factors of bound nucleons. The parametrization should also be applicable to the transverse cross section in  neutrino scattering. If the transverse enhancement  originates from  meson exchange currents (MEC), then it is theoretically expected that any enhancement in the  longitudinal or axial contributions is small.    We present the predictions of the "Transverse Enhancement" model (which is based on electron scattering data only)  for the   $\nu_\mu, \bar{\nu}_\mu$ differential and total QE cross sections for nucleons bound in carbon. The $Q^2$ dependence of the transverse enhancement is observed to resolve much of the  long standing discrepancy ({\em "Axial Mass Anomaly"}) in the QE total cross sections and differential distributions between
  low energy and high energy neutrino experiments on nuclear targets.
\end{abstract}

\maketitle

%%%%%%%%%%%%%%%%%%%%%%%%%%%%%%%%%%%%%%%%%%%%
%% MAINMATTER
%%%%%%%%%%%%%%%%%%%%%%%%%%%%%%%%%%%%%%%%%%%%

%\section{Introduction}

  Models which assume that $\nu_{\mu},\bar{\nu}_\mu$ quasielastic (QE) scattering on nuclear targets can be described in terms of scattering from independent nucleons bound in a nuclear potential (e.g. Fermi gas or spectral functions) 
do not provide an adequate representation of measured differential and total 
QE cross sections for low energy ($\approx 1 GeV$) $\nu_\mu$ scattering on nucleons bound in  carbon (MiniBooNE) 
 and oxygen (K2K and T2K). The measured QE total cross sections are 20-30\% larger than the model and 
 the  differential distributions in $Q^2$ are also inconsistent. 
%and Iron\cite{MINOS}. 
The vector and axial form factors that are used in  "independent nucleon" models 
are the free nucleon form factors extracted from electron and $\nu_{\mu},\bar{\nu}_\mu$ scattering data on hydrogen (H)  and
deuterium (D).

The disagreement between the measured low energy $\nu_\mu$  differential and total QE cross sections 
on nuclear targets and the predictions of the "independent nucleon" model has been attributed to an incomplete description of nuclear effects. 
These additional nuclear effects have been parametrized as an ad-hoc change in the 
the axial form factor mass parameter from the value measured for free nucleons $M_A^{free}=1.014\pm0.014 ~GeV$ to $M_A^{eff} \approx 1.30  ~GeV$.

However, at  high neutrino energies, the measured total and differential QE cross sections on nuclear targets are consistent with  models which assume that the scattering is on independent nucleons with free nucleon form factors.  For example,   $M_A$ of  $0.979 \pm 0.016$ GeV has been extracted  by Kuzmin {\em et al.}  from a global analysis
%\cite{quasinuclear}
of the differential distributions and total QE cross sections  measured in  $all$ high  energy $\nu_{\mu},\bar{\nu}_\mu$ experiments on nuclear targets. 
This discrepancy between low energy and high energy data has sometimes been referred to as the {\em axial mass anomaly}.

 \begin{figure}[h]
\includegraphics[width=3.3 in,height=2.4in]{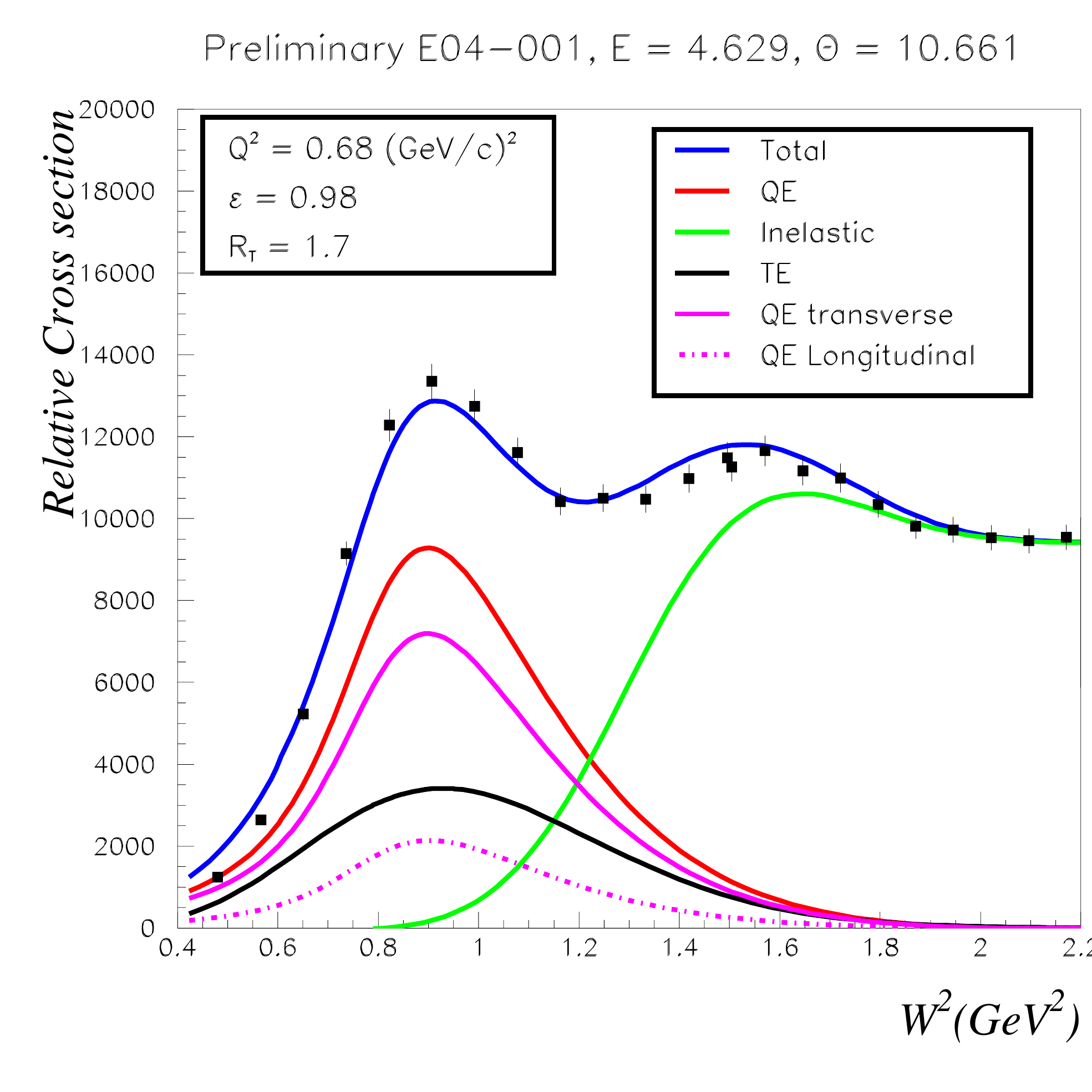}
\includegraphics[width=3.3in,height=2.5in]{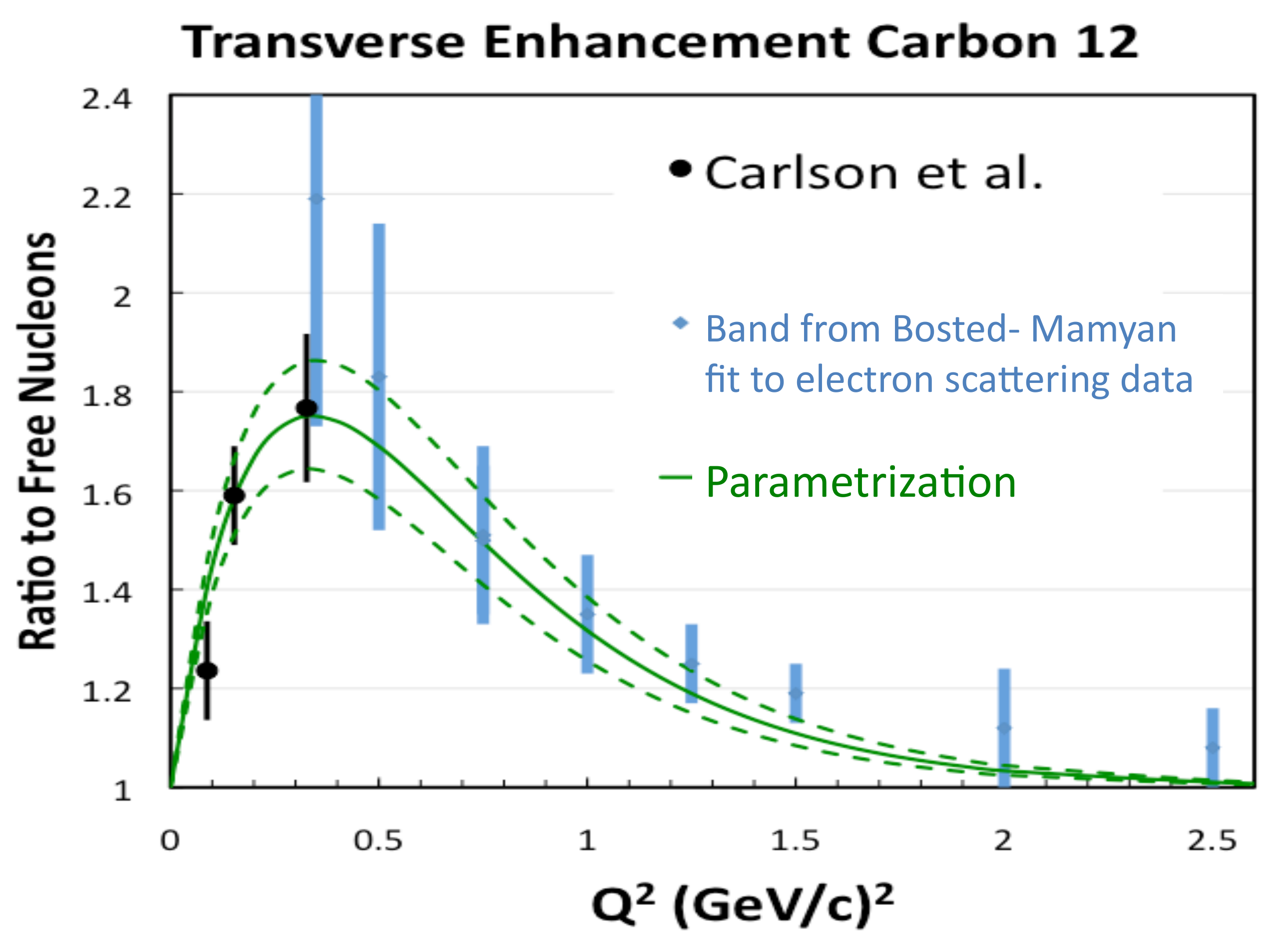}
\caption{ {\bf Left}: Example of the fit  to preliminary electron scattering data from the JUPITER collaboration (Jefferson Lab experiment E04-001) on a carbon target. Shown are the contributions from the transverse QE (solid pink), longitudinal QE (dashed pink), total QE (solid red), inelastic pion production processes (solid green), and the  transverse excess (TE) contribution (solid black line).
Here, $Q^2=$~0.68 $GeV/c^2$ at the QE peak. 
 {\bf Right}:  The transverse enhancement ratio (${\cal R}_{T}$) as a function of $Q^2$. Here, ${\cal R}_{T}$ is  ratio of the  integrated transverse response function
for QE  electron scattering on nucleons bound in  carbon divided by 
the integrated response function for independent nucleons. 
 The black points are extracted from Carlson $et~al$, and
the blue bands are extracted from the fit  to QE data from the JUPITER collaboration. The curve is a fit to ${\cal R}_{T}(Q^2)$ of the form  ${\cal R}_{T}=1+AQ^2e^{-Q^2/B}$,  with $A=6.0$ and $B=0.34~(GeV/c)^2$. The dashed lines are estimated upper and lower error bands.  }
\label{GMPN}
\end{figure}

Studies of QE  electron  scattering on nuclear targets indicate
that only the longitudinal part of the QE cross section  can be described in terms of a universal response function of 
independent nucleons bound in a nuclear potential (and free nucleon form factors).  In contrast, a significant additional enhancement with respect to the model is observed in the transverse part of the QE cross section.

The enhancement  in the transverse QE cross section has been attributed to meson exchange currents (MEC) in a nucleus. 
In MEC models  the enhancement is primarily in the transverse part of the QE cross section, while the  enhancement in the longitudinal QE cross section is small.  The conserved vector current hypothesis (CVC) implies that  the  corresponding  vector structure functions for  the QE cross sections in $\nu_{\mu},\bar{\nu}_\mu$ scattering can be expressed in terms of the  structure functions measured in electron scattering on nuclear targets. Therefore,
 there should also be a vector transverse  enhancement in $\nu_{\mu},\bar{\nu}_\mu$ scattering. In addition, in  meson exchange currents  models, the enhancement in the 
axial part of the $\nu_{\mu},\bar{\nu}_\mu$ QE cross section on nuclear targets is also small. Therefore, the  axial form factor for bound nucleons is expected to be the same as the axial form factor for free nucleons.

The longitudinal and transverse response  functions for QE scattering
on different nuclei (A=12 ,40 and 56) were   extracted by Donnely $et~al$ for $Q^2$ values of 0.09, 0.15, and 0.33 $(GeV/c)^2$. Carlson {\em et al.} use these measured longitudinal and transverse response functions  to extract ${\cal R}_{T}$, which
the ratio of the  integrated transverse and the integrated longitudinal response functions (assuming free nucleon form factors).  Since the universal longitudinal  response function can be described by a model of independent nucleons bound in a nuclear potential,  ${\cal R}_{T}$ is equivalent to the ratio of the transverse cross sections of  bound and free nucleons.
  
 % The technique of using the ratio of  longitudinal and transverse QE structure functions
%to determine the transverse enhancement in the response functions for QE scattering %is less reliable for $Q^2>0.5~(GeV/c)^2$, because at  high values of $Q^2$ the  %longitudinal contribution to the QE cross section is small. 

We extract the transverse enhancement at higher values of $Q^2$ from a fit to  
existing electron scattering data on nuclei and  preliminary data from the JUPITER 
collaboration  (Jefferson lab experiment E04-001).  The fit (developed 
by P. Bosted and V. Mamyan) provides a description of inclusive electron scattering cross sections on a range of nuclei with $A > 2$.  An example of the fit for a carbon spectrum is shown on the left panel of Fig.\ref{GMPN}.

The Bosted-Mamyan inclusive fit is a sum of four components: 
\begin{itemize}
\item The longitudinal  QE contribution extracted from H and D experiments (smeared by Fermi motion in carbon)
\item  The transverse  QE contribution extracted from H and D experiments (smeared by Fermi motion in carbon)
\item  The contribution of inelastic pion production processes from H and D (smeared by Fermi motion in carbon).
\item  A transverse excess (TE) contribution (determined by the fit)
\end{itemize}

The right panel of Fig.~\ref{GMPN} shows the values of ${\cal R}_{T}$  as a function of $Q^2$.  The black points are extracted from Carlson $et~al$, and
the higher $Q^2$ blue bands are from the fit to the  QE data from the JUPITER collaboration.  The data are parametrized by the expression:
  ${\cal R}_{T}=1+AQ^2e^{-Q^2/B}$
   with $A=6.0$ and $B=0.34~(GeV/c)^2$.  The electron scattering data indicate that the transverse enhancement is maximal near $Q^2=0.3~(GeV/c)^2$ and is small for $Q^2$ greater
than $1.5~(GeV/c)^2$.  The dashed lines are the estimated upper and lower error bands 

\begin{figure}
\includegraphics[width=3.503in,height=2.1in]{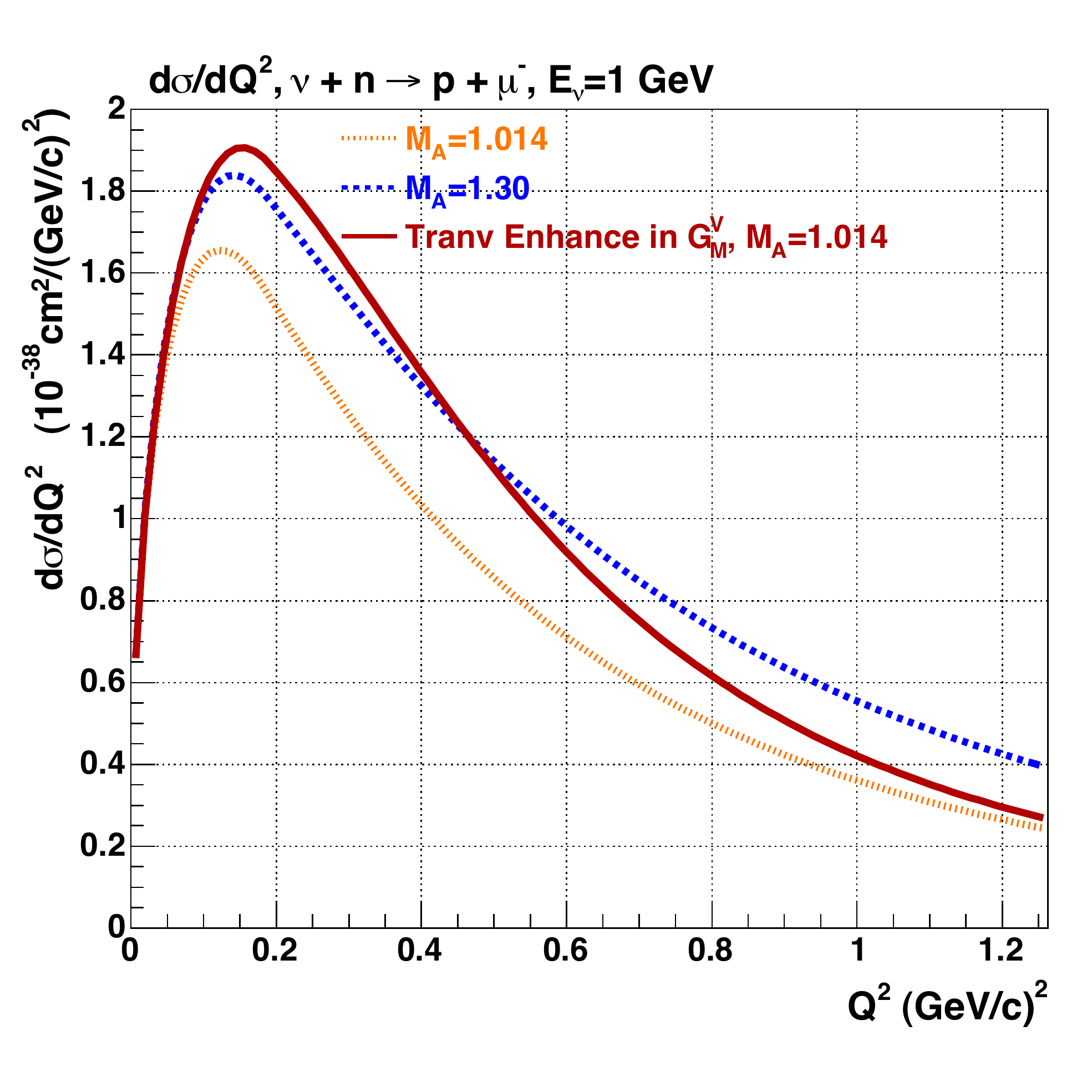}
\includegraphics[width=3.503 in,height=2.1in]{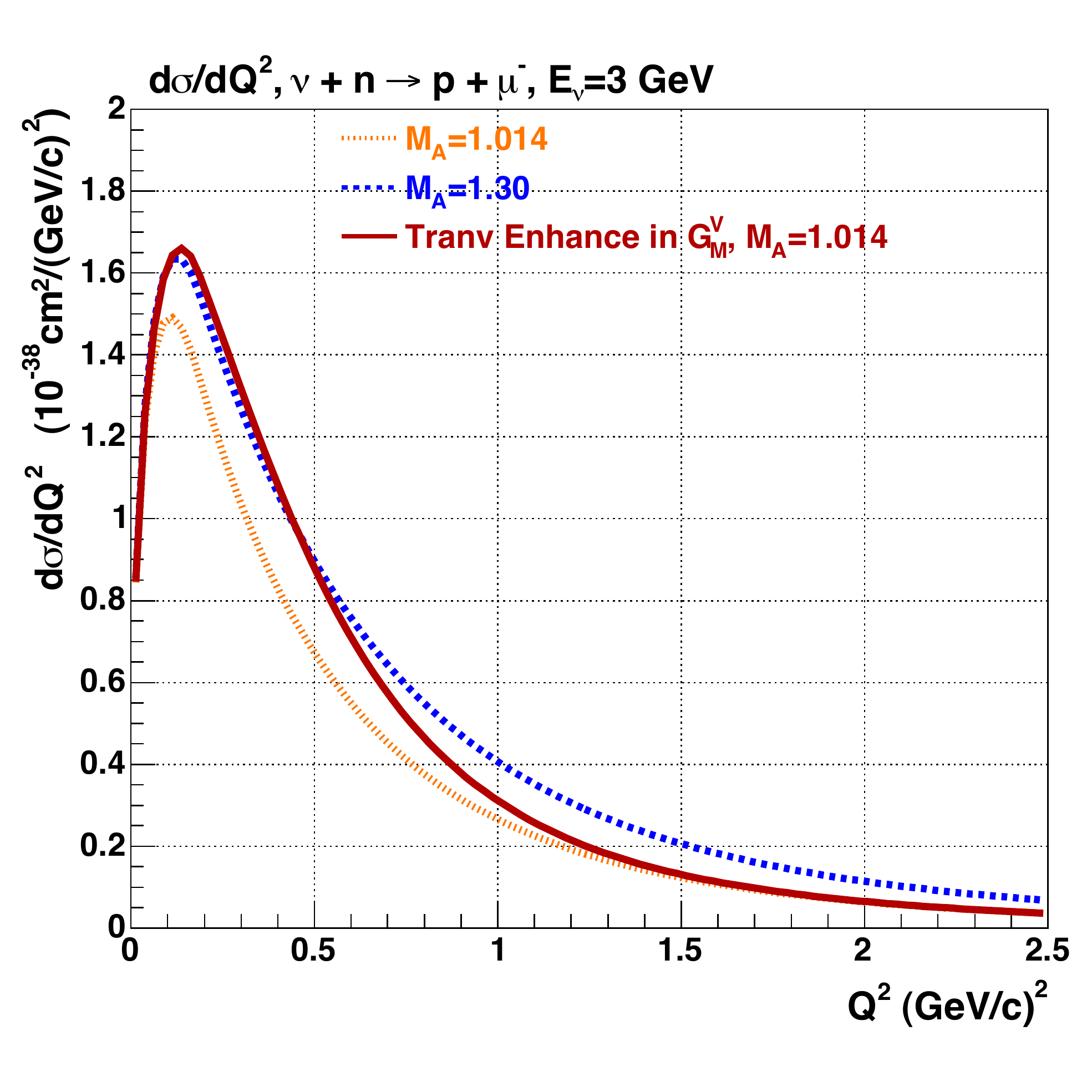}
\caption{The $\nu_{\mu}$ QE differential cross section (d$\sigma$/d$Q^2$) 
for nucleons bound in carbon 
 as a function of $Q^2$. Shown are the  prediction of the  "Independent Nucleon" model with $M_A$=1.014 GeV
(orange dotted line), and with  $M_A$=1.3 GeV (blue dashed  line). The red line is the   "Independent Nucleon" model  with  $M_A$=1.014 GeV and including "Transverse Enhancement".  {\bf Left panel}:  $E_\nu$ =1  GeV (maximum accessible $Q^2_{max} = 1.3~ (GeV/c)^2$). {\bf Right panel}:  $E_\nu$ = 3 GeV  ($Q^2_{max} = 4.9~ (GeV/c)^2$). 
 }
\label{diff1}
\end{figure}

 Fig. \ref{diff1} shows  $d\sigma$/d$Q^2$ predictions  for  $\nu_{\mu}$ QE scatterring  
on  carbon 
 as a function of $Q^2$. Shown are predictions of the  "Independent Nucleon" model with $M_A$=1.014 GeV  (orange dotted line),  with $M_A$= 1.3 GeV (blue dashed  line),  and with $M_A$=1.014 GeV including "Transverse Enhancement" (red line).
  The left panel is for  $E_\nu$ =1  GeV  and the right panel is for $E_\nu$ = 3 GeV.

  For $Q^2<0.6~(GeV/c)^2$ the predictions for d$\sigma$/d$Q^2$ with  $M_A$=1.014 GeV and including "Transverse Enhancement" are similar to d$\sigma$/d$Q^2$ with  $M_A$=1.3 GeV. 
 The  maximum accessible  $Q^2$ for 1 GeV neutrinos is $1.3~(GeV/c)^2$. Therefore,  fits to  d$\sigma$/d$Q^2$ for $E_\nu$ =1  GeV (e.g. MiniBooNE)  would yield $M_A \approx 1.2~GeV$.  
 
 In the high $Q^2$ region  ($Q^2>1.2~(GeV/c)^2$), the magnitude of the  
  "Transverse Enhancement" is small.  The maximum accessible  $Q^2$ for 3 GeV neutrinos is $4.9~(GeV/c)^2$.  In order to reduce the sensitivity to modeling of  Pauli blocking,  experiments at higher energy   typically remove the
  lower $Q^2$ points in fits for $M_A$.   Consequently,  fits
to d$\sigma$/d$Q^2$ measured in high energy experiments would yield a value of 
 $M_A$ which is smaller than $1.014$ GeV  because for $Q^2>0.5~(GeV/c)^2$ the slope of d$\sigma$/d$Q^2$ in the transition region between low and high $Q^2$ 
 is steeper than for $M_A$=1.014 GeV. This is consistent  with the fact that the average 
 $M_A$ extracted from  high energy neutrino experiments on nuclear targets by Kuzmin {\em et al.} is  $0.979 \pm 0.016$.
 
\begin{figure}
\includegraphics[width=3.503in,height=2.5in]{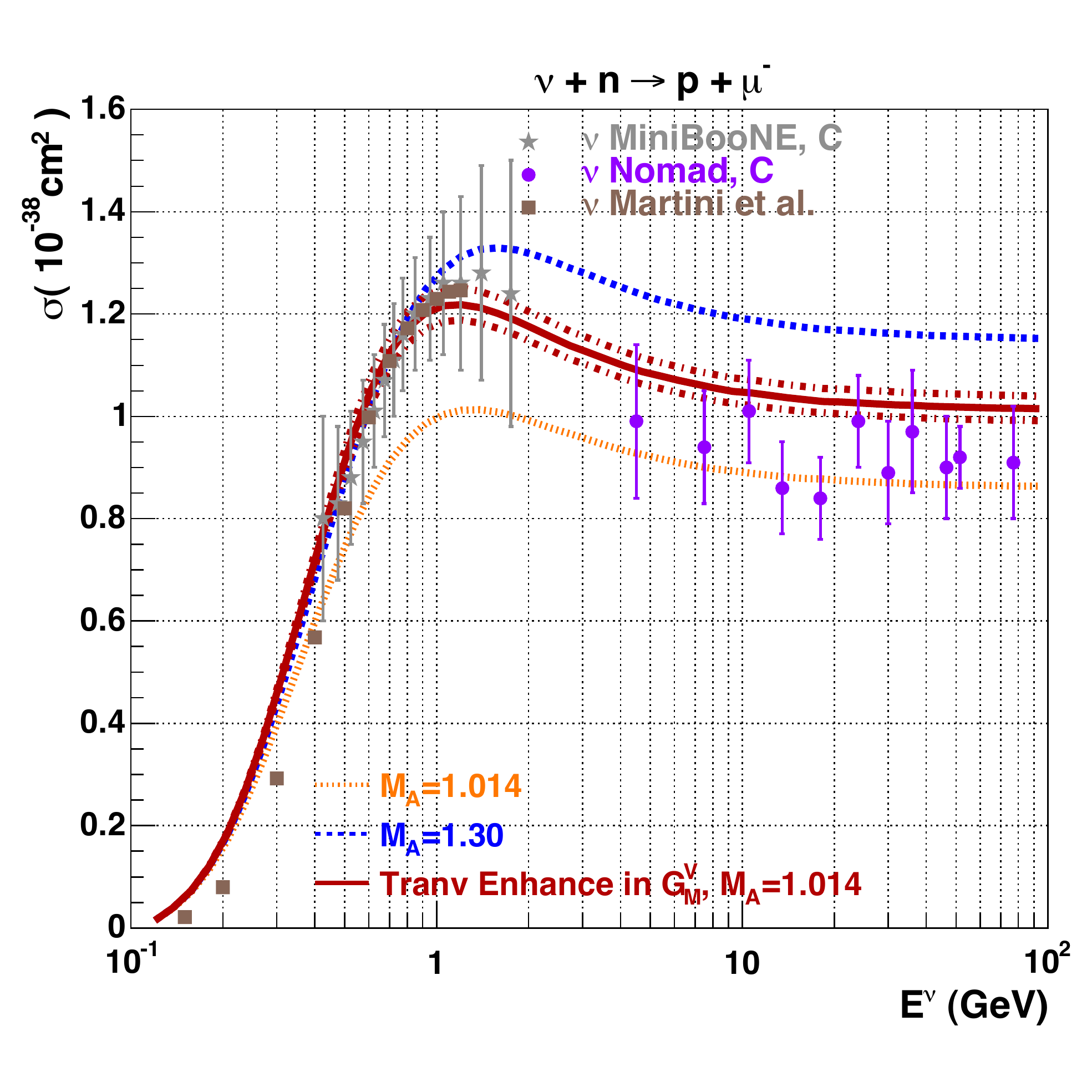}
\includegraphics[width=3.503 in,height=2.5in]{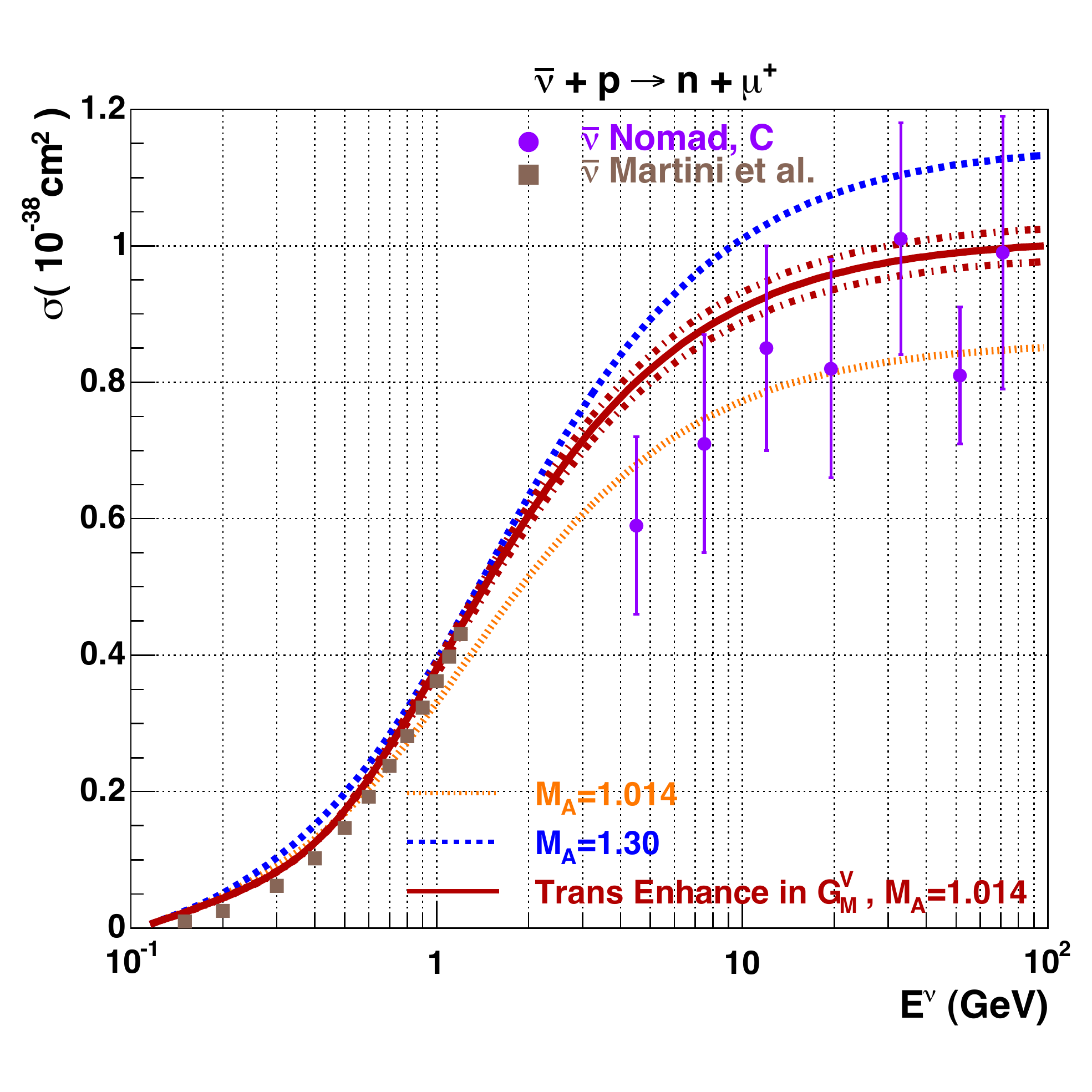}
\caption{Comparison of predictions for the   $\nu_{\mu}$ (left) and $\bar{\nu}_\mu$(right)   total QE cross section section as a function of $E_\nu$ for 
the "Independent Nucleon" model for $M_A$=1.014 GeV, $M_A$=1.3 GeV and $M_A$=1.014 GeV with "Transverse Enhancement" and  (shown with error bands). Also shown are predictions of  
  "QE+np-nh RPA"  MEC model of Martini {\em et al.}. (Predictions for this model have only been published for $E_\nu~< 1.2$ GeV). 
  The data points are measurements of MiniBooNE  and NOMAD.
  }
  \label{cross}
\end{figure}

Fig. \ref{cross} shows the predictions for the   $\nu_{\mu}$ (left) and $\bar{\nu}_\mu$(right)   total QE cross section as a function of energy for 
the "Independent Nucleon" model  with $M_A$=1.014 GeV, and with $M_A$=1.3 GeV , compared 
to the  "Independent Nucleon" model with  $M_A$ =1.014 GeV and   "Transverse Enhancement" (with error bands). Also shown are predictions of the  
  "QE+np-nh RPA"  Meson Exchange Current model of Martini {\em et al.}. (Predictions for this model have only been published for $E_\nu$~<~1.2 GeV). 
  The data points are the measurements of MiniBooNE  and NOMAD.
  
 The predictions of the   "Independent Nucleon" model with  $M_A$ =1.014 GeV  including   "Transverse Enhancement"  are  in  agreement with the MiniBooNE  cross sections 
 at low energies, and are also  consistent with the NOMAD  cross section measurements at high energies (within experimental errors), thus resolving the {\em"axial mass anomaly"}.
 
 Additional details and a list of references discussed in this paper can be found in   A. Bodek, H. S. Budd, M. E. Christy,
{\em  "Neutrino Quasielastic Scattering on Nuclear Targets: Parametrizing Transverse Enhancement (Meson Exchange Currents)" }
{\bf  arXiv:1106.0340} [hep-ph], and references therein.

%%%%%%%%%%%%%%%%%%%%%%%%%%%%%%%%%%%%%%%%%%%%%%%%
%% BACKMATTER
%%%%%%%%%%%%%%%%%%%%%%%%%%%%%%%%%%%%%%%%%%%%%%%%
%%%%%%%%%%%%%%%%%%%%%%%%%%%%%%%%%%%%%%%%%%%%%%%%
%% The bibliography can be prepared using the BibTeX program or
%% manually.
%%
%% The code below assumes that BibTeX is used.  If the bibliography is
%% produced without BibTeX comment out the following lines and see the
%% aipguide.pdf for further information.
%%
%% For your convenience a manually coded example is appended
%% after the \end{document}
%%%%%%%%%%%%%%%%%%%%%%%%%%%%%%%%%%%%%%%%%%%%%%%%

%%%%%%%%%%%%%%%%%%%%%%%%%%%%%%%%%%%%%%%%%%%%%%%%
%% You may have to change the BibTeX style below, depending on your
%% setup or preferences.
%%
%%
%% For The AIP proceedings layouts use either
%%%%%%%%%%%%%%%%%%%%%%%%%%%%%%%%%%%%%%%%%%%%

\bibliographystyle{aipproc}   % if natbib is available
%\bibliographystyle{aipprocl} % if natbib is missing

%%%%%%%%%%%%%%%%%%%%%%%%%%%%%%%%%%%%%%%%%%%
%% You probably want to use your own bibtex database here
%%%%%%%%%%%%%%%%%%%%%%%%%%%%%%%%%%%%%%%%%%%
\bibliography{sample}

%%%%%%%%%%%%%%%%%%%%%%%%%%%%%%%%%%%%%%%%%%%
%% Just a reminder that you may have to run bibtex
%% All of it up to \end{document} can be removed
%% if you don't like the warning.
%%%%%%%%%%%%%%%%%%%%%%%%%%%%%%%%%%%%%%%%%%%
\IfFileExists{\jobname.bbl}{}
 {\typeout{}
  \typeout{******************************************}
  \typeout{** Please run "bibtex \jobname" to optain}
  \typeout{** the bibliography and then re-run LaTeX}
  \typeout{** twice to fix the references!}
  \typeout{******************************************}
  \typeout{}
 }

\end{document}